\documentclass{article}

\usepackage{graphicx}
\usepackage{latexsym}
\usepackage{amsfonts,amsmath,amssymb}
\usepackage[utf8]{inputenc}
\usepackage[english]{babel}
\usepackage{color}
\usepackage[hidelinks]{hyperref}
\hypersetup{colorlinks=false}
\usepackage[a4paper,margin={2.5cm,2.5cm,2.5cm,2.5cm}]{geometry}
\usepackage{url}
\usepackage{authblk}
\usepackage[compress,numbers]{natbib}

\title{Race, Religion and the City:
Twitter Word Frequency Patterns Reveal Dominant Demographic Dimensions in the United States}
\author[1]{Eszter Bok{\'{a}}nyi$^{*}$}
\author[1,2]{Dániel Kondor}
\author[1]{L{\'{a}}szl{\'{o}} Dobos}
\author[1]{Tam{\'{a}}s Seb{\H{o}}k}
\author[1]{J{\'o}zsef St{\'{e}}ger}
\author[1]{Istv{\'{a}}n Csabai}
\author[1]{G{\'{a}}bor Vattay}
\affil[1]{Department of Physics of Complex Systems, E{\"{o}}tv{\"{o}}s Lor{\'{a}}nd University, Pázmány Péter sétány 1/A, H-1117 Budapest, Hungary}	
\affil[2]{SENSEable City Laboratory, Massachusetts Institute of Technology, 77 Massachusetts Avenue, Cambridge, MA 02139, USA}	
\date{\small \textit{$^*$bokanyi@complex.elte.hu}}
\begin{document}

\maketitle

\begin{abstract}

Recently, numerous approaches have emerged in the social sciences to exploit the opportunities made possible by the vast amounts of data generated by online social networks (OSNs). Having access to information about users on such a scale opens up a range of possibilities -- from predicting individuals' demographics and health status to their beliefs and political opinions -- all without the limitations  associated with often slow and expensive paper-based polls. A question that remains to be satisfactorily addressed, however, is how demography is represented in the OSN content -- that is, what are the relevant aspects that constitute detectable large-scale patterns in language? Here, we study language use in the US using a corpus of text compiled from over half a billion geo-tagged messages from the online microblogging platform Twitter. Our intention is to reveal the most important spatial patterns in language use in an unsupervised manner and relate them to demographics. Our approach is based on Latent Semantic Analysis (LSA) augmented with the Robust Principal Component Analysis (RPCA) methodology, which permits identification of the data's main sources of variation with an automatic filtering of noise and outliers without influencing results by a priori assumptions. We find spatially correlated patterns that can be interpreted based on the words associated with them. The main language features can be related to slang use, urbanization, travel, religion and ethnicity, the patterns of which are shown to correlate plausibly with traditional census data. Apart from the standard measure of linear correlation, some relations seem to be better explained by boolean implications, suggesting a threshold-like behavior where demographic variables influence the users' word use. Our findings thus validate the concept of demography being represented in OSN language use and show that the traits observed are inherently present in the word frequencies without any previous assumptions about the dataset. Thus, they could form the basis of further research focusing on the evaluation of demographic data estimation from other big data sources, or on the dynamical processes that result in the patterns found here.

\end{abstract}

\section{Introduction}

Geography plays an important role in many social phenomena: clearly, many aspects of life are influenced by the possibilities offered by the environment in which one lives \cite{Brain2005,Quillian1999,Sampson2009,Iceland2006,Bruch2006,Bettencourt2007}. As such, uncovering the spatial structures and the dynamics of changes in them has for some time been a focus of the scientific community and policymakers. In line with this, governments and local authorities invest significant resources in creating and maintaining databases of census data, including several variables describing the local population and economic activity on the regional scale. These data-collection and monitoring activities are usually limited by the significant efforts required to obtain and process data, prompting researchers and professionals to look for alternative data sources and methods that can complement traditional data collection and which could be integrated with modeling and research efforts \cite{Cummings2012,OConnor2010,Deville,Botta2015,Louail,Frias,Ratti2007cellularcensus}.

In the past two decades, there has been a significant growth in the amount of data collected about individuals that has been made available for research purposes. This has had a large impact on social science research where empirical studies were previously limited by the cost and effort associated with data collection. This includes studies focusing on how modern data collection methods can be used to reveal the spatial structure in society on several scales, and how quantities measured in the online or abstract environments are connected to real-world phenomena. Two common data sources are mobile phone networks, where user activity and aggregated measures of network utilization are recorded at the antenna level as part of regular operation \cite{Krings} and online social networks (OSNs) \cite{mislove}, where the content publicly shared by users in many cases includes their position \cite{Cheng2011}. Some other data sources with promising application possibilities include monetary transactions \cite{george,george2,bbva}, GPS traces from cars \cite{cargps,returners} and other devices and public transportation usage as recorded by electronic payment systems \cite{oyster,oyster2}.

Using these data, previous research has shown that it is possible to obtain accurate and up-to-date measures of population density \cite{Deville} or crowd size at sports events or in airports \cite{Botta2015}. Furthermore, the demographic features of a city or a country can be estimated by parsing OSN user names or user profile descriptions \cite{Sloan2015,Longley2015}. By focusing on the community structure instead of estimating features of individuals, networks of connections among mobile phone or social network users reveal geographic clustering on large scales \cite{george2,delineating,spatialfingerprints}, Twitter users' language choice reflects different cultural communities \cite{Mocanu2013}, while user activity has been used on urban scales as an innovative method of land use detection \cite{Ratti2007cellularcensus,Grauwin,Frias,Louail}. In addition to land use data, commuting and mobility patterns in the city \cite{Gonzalez2008mobility,Jiang} and larger scale travel trends can also be investigated with the help of mobile and OSN networks \cite{gowalla,Hawelka2014,Simini}.

Apart from looking at the spatio-temporal patterns, analysing the content of users posted in OSNs can provide further insights, adapting text mining methods and results which have been previously developed and obtained on the growing corpus of digital texts \cite{Landauer1997,Deerwester1990,Schwartz2013a,Petersen2012,Perc2012}. From predicting heart-disease rates of an area based on its language use \cite{Eichstaedt2015}, connecting health measures to photo scenicness ratings \cite{Seresinhe2015} or relating unemployment to social media content \cite{unemployment,Pavlicek2015} to forecasting stock market moves from search semantics \cite{Curme2014}, many studies have attempted to connect online media language and metadata to real-world outcomes. Various studies have analyzed spatial variation in the OSN messages' texts and its applicability to several different questions, including user localization based on the content of their posts \cite{Cheng2010,Backstrom2010}, empirical analysis of the geographic diffusion of novel words, phrases, trends and topics of interest \cite{travelingtrends,Eisenstein2012}, measuring public mood \cite{Mitchell2013}.

In these studies, either a priori models were used, or a model was built with a supervised learning method, with a focus on the specific phenomenon, meaning the exploitation of only one aspect (user name, user profile description, misspelled words, words connected to fatigue etc.), yet possibly neglecting the dataset's other features. While being effective, there remain the following questions: (a) what are main patterns in the data in general; (b) can they be discovered without making a priori assumptions about what to look for; (c) can we relate these patterns to relevant real social phenomena.

In this study our goal is to analyze in an unsupervised manner how and to what extent regional-scale demographic attributes are represented in social media posts. We approach this using geo-tagged short messages (tweets) posted on the Twitter microblogging service as a source of large-scale digital corpus. We employ a combination of Latent Semantic Analysis (LSA) \cite{Deerwester1990} and Robust Principal Component Analysis (RPCA) \cite{Lin2010,Candes2011}, which permits us the automated identification of the most significant topics and language use features with regional variation on Twitter. We use tweets posted in the USA over a 3-year period aggregated at the county-level. This allows comparison with census data at the same level, thus allowing us to draw some hypotheses about the driving forces behind regional language dissimilarity patterns.

\section{Methods}
\subsection{Twitter dataset}
We use the datastream freely provided by Twitter through their Application Program Interface (API), which amounts to approximately 1\% of all sent messages. In this study, we focus on the part of the datastream with geolocation information. These geolocated tweets originate from users who chose to allow their mobile phones to post the GPS coordinates along with a Twitter message. The total geolocated content was found to only comprise a small percentage of all tweets; therefore with data collection focusing only on these, a large fraction of all geo-tagged tweets can be gained \cite{Morstatter2013}.

Our dataset includes a total of 335 million tweets from the contiguous United Stated of America collected between February 2012 and June 2013. These are all geotagged – that is, they have GPS coordinates associated with them. We construct a geographically indexed database of these tweets, permitting the efficient analysis of regional features \cite{Dobos2013}. Using the Hierarchical Triangular Mesh (HTM) scheme for practical geographic indexing \cite{Szalay2007,geomhtm}, we assigned a US county to each tweet. County borders are obtained from the GAdm database \footnote{\url{http://gadm.org}}.

\subsection{Latent Semantic Indexing and Robust Principal Component Analysis}

We aim to use a type of vector space model on our Twitter corpus, where documents correspond to county-level aggregated tweets. The terms we consider are raw words obtained after a tokenization process, - that is, we apply a ’word-bag’ approach to our documents, effectively limiting any analysis to word frequencies and ignoring relations among words and longer phrases. We filter stop-words in several languages (most important being English and Spanish) to remove most common but uninformative terms from our data.

We construct a term-document matrix $W_{ij}$ as the as the number of occurrences of the $i$-th word in the $j$-th cell. As the population density of the USA is very heterogeneous, the number of word occurrences in each county is also heterogeneous. To improve the quality of the dataset, we only include counties t contain at least 10000 occurrences of at least 500 individual words. We also limit the words used to those with at least 10000 occurrences in at least 1000 individual counties. This way there remain 2800 counties and 10132 words, which form the $W_{ij}$ word occurrence matrix. We normalize $W_{ij}$ so that the elements are the relative frequencies of words in each county: $X_{ij} \equiv W_{ij}/\sum_k W_{kj}$, i.e.~we normalize each element by the total number of words posted in that county; this is called inverse document frequency weighing in text-mining literature.

To identify all possible regional characteristics of language usage, we rely on techniques known from the field of natural language processing. There exist many feature or topic extraction methods, all of them aiming to reduce the dimensionality of the data by finding related or similar words and documents. A common approach is Latent Semantic Analysis (LSA) {\cite{Deerwester1990,Gotoh1997}}, which applies Singular Vector Decomposition (SVD) on a word by document matrix derived from the corpus. This method groups words together based on their semantic similarity {\cite{Landauer1997}}, creating 'feature' documents, of which the first few represent the concepts causing the most variation in the data. A notable achievement of LSA is that it is an unsupervised learning method, thus providing information about the corpus without using a priori assumptions or any arbitrary preselections based on the purpose of the examination. 

According to the nature of our dataset, there are several users who generate automated messages like weather stations, advertisers or tornado and earthquake advisories, which are considered as noise in our investigations. Especially in sparsely inhabited areas, these outlier messages can account for a large fraction of the dataset. Also, highly localized features, such as~tourist attractions, can generate outliers of significant volume. This can result in highly localized outliers dominating the results of the SVD, making identifying relevant structure challenging. Applying the Robust PCA method {\cite{Candes2011,Lin2010}} allows us to preprocess the matrix before further analysis by separating it into a low-rank and a sparse part, whose principal components can then be computed and analyzed separately. This means that the original data matrix is written as a sum of two parts:
\begin{equation}
	X = X^{\textrm{S}} + X^{\textrm{LR}} \, \textrm{,}
\end{equation}
where $X^{\textrm{S}}$ is a sparse matrix and $X^{\textrm{LR}}$ contains the dense but low-rank part of the data. The mathematical condition for finding $X^{\textrm{S}}$ and $X^{\textrm{LR}}$ is minimizing the sum 
\begin{equation}
	\lambda \| X^{\textrm{S}} \|_1 + \| X^{\textrm{LR}} \|_{\sigma} \, \textrm{,}
\end{equation}
where for a matrix $X$ of dimensions $n_1 \times n_2$ with $n_1 \geq n_2$, $\lambda \equiv 1 / \sqrt{n_1}$, and the norms are the $l_1$ and nuclear norms respectively:
\begin{gather}
\begin{split}
	\| X \|_1 = \sum_{ij} |X_{ij}| \\
	\| X \|_{\sigma} = \sum_i \sigma_i (X) \, \textrm{.}
\end{split}
\end{gather}
Here $\sigma_i (X)$ denotes the $i$-th singular value of $X$. An efficient algorithm for finding $X^{\textrm{S}}$ and $X^{\textrm{LR}}$ is the inexact augmented Lagrangian method~{\cite{Lin2010}} (Matlab code developed by the authors of~{\cite{Lin2010}} implementing the algorithm is publicly available \footnote{\url{http://perception.csl.illinois.edu/matrix-rank/sample_code.html}}). Employing this method results in the sparse part containing most of the outliers, and and in true language use variations to be represented in the low-rank part. Due to the structure of our data matrix, and the employed Robust PCA method, we choose not to subtract averages from the data; of course, this will probably result in average word frequencies dominating the first principal component. We further analyze only the results of the LSA of the low-rank component.

\subsection{Demographic data}

To discover possible governing factors of the geographical language variation patterns and connections between topics and their geography, we correlate right singular vectors with a variety of demographic data series from the 2010 US Census \footnote{\url{http://www2.census.gov/census_2010/}, \url{http://www.census.gov/support/USACdataDownloads.html}}, 2011 American Community Survey (ACS) 5 year estimates concerning educational attainment by counties \footnote{\url{http://www.census.gov/programs-surveys/acs/}}, county business patterns according to North American Industry Classification System (NAICS) classification \footnote{\url{http://www.census.gov/econ/cbp/}} and church adherence rates and congregations numbers per county provided by the the Association of Religion Data Archives (ARDA) \footnote{\url{http://www.thearda.com}}.

\subsection{Boolean relationship detection}

Apart from evaluating linear correlation measures with the singular vectors, we also carry out a boolean relationship detection, using the methodology of Sahoo et al. \cite{Sahoo_2008}, which is based on calculating a test statistic based on the contingency table of the scatterplots (e.g.~Fig.~\ref{fig:sdemo}f-j, see the next section for an interpretation of the results displayed) after creating the four segments of the data with a horizontal and a vertical limit. We find the most significantly sparse segment by setting the limits so that the test statistic gives a maximum for the specific segment. During the calculations, we set an error bar on both side of the limits, and points being in this error zone are not taken into consideration when testing for the sparseness.

If the contingency table is

\begin{table}[h]

	\centering
	\begin{tabular}{c|cc|c}
			& $A$ low & $A$ high & $\Sigma$\\\hline
			$B$ low & $m_{00}$ & $m_{01}$ & $b_0$\\
			$B$ high & $m_{10}$ & $m_{11}$ & $b_1$\\\hline
			$\Sigma$ & $a_0$ & $a_1$ & $s$
	\end{tabular}
\end{table}

The test statistic for the four segments is
\[\delta=\frac{m_{ij}-\langle m_{ij} \rangle}{\sqrt{\langle m_{ij} \rangle}},\]
where $\langle m_{ij} \rangle$ denotes the expected value in case of independent variables
\[\langle m_{ij}\rangle=\frac{a_i}{s}\frac{b_j}{s}\cdot s.\]

If there are some points left in the segment, they are considered as an error, and the measure of error would be
\[\epsilon=\frac{1}{2}\left(\frac{m_{ij}}{m_{i0}-m_{i1}}+\frac{m_{ij}}{a_i}\right).\]

We consider a segment significantly sparse if $\delta>3$, and $\epsilon<0.2$.

Then in the whole range of variables $A$ and $B$ (using 100 steps in both directions and an error boundary of 1,5\% for the skipping of points near the borders) we measure $\delta$ and $\epsilon$ values, and take the segmentation with the maximum $\delta$ for the sparse areas, where $\epsilon$ is still low enough.

\section{Results}

Using a corpus of over 335 million geo-tagged tweets posted in the USA, we compile word-frequency distributions for each US county, and then apply the automatic filtering and feature selection method described. We analyze the features found with this technique by considering the connection between geographic and semantic distances (Fig. 1.), and by plotting right singular vectors on the map (Fig. 2a-e.) and displaying left singular vectors as wordclouds (positive weights Fig. 2f-j., negative weights Fig. 2k-o.).

First, we find that the method applied successfully uncovers some coherent topics, especially in the first few singular vectors, where singular values are still great enough for the topic to give a significant variance of the dataset. As we deliberately choose not to subtract averages from the $X_{ij}$ matrix, the first component shows no discernible pattern, and corresponds to the most common words in the sample. From the second singular vector, however, one or both ends (it can be either negative or positive, as singular vectors can arbitrarily be multiplicated by a minus sign) of each of the most important semantic features on the wordclouds can be related to a certain language style, concept or lifestyle.

The words giving the largest contribution to the pattern of the second left singular vector (Fig.~\ref{fig:svect}f.) mark a strong presence of slang in the sample. This includes forms with alternate spelling like 'aint', 'gotta'; swearing like 'ass', 'hoe', 'bitch'; abbreviations of common phrases like 'tryna', 'imma', 'kno', 'yall'; OSN-specific slang such as 'oomf' which stands for 'one of my followers' (i.e.~on Twitter); a very specific misspelling of 'goodmorning' (instead of 'good morning'); and variations of the racial slurs 'nigga' and 'niggas'. Swear words and abbreviations typical for online language also dominate this end of the component. The next most important feature, which can be found in the third vector (Fig.~\ref{fig:svect}l.), identifies words connected to urban lifestyle like eating out ('pizza', 'grill'), drinking coffee ('coffee', 'cafe', 'starbucks'), education ('university', 'library', 'campus') or working out ('gym', 'fitness').

Further dominating concepts are travel ('enjoying', 'trip', 'pic', 'hotel') in the fourth singular vector (Fig.~\ref{fig:svect}h.)  and religion ('lord', 'prayers', 'praying', 'blessed') alongside with positive content ('glad', 'thankful', 'wonderful', 'proud') in the negatively weighed words of the fifth singular vector (Fig.~\ref{fig:svect}n.). In this case, the opposite end can also be easily interpreted: the faith-related words in the fifth component are countered by an increased usage of profanity present among words with positive weights (Fig.~\ref{fig:svect}i.). This might be the consequence of people tweeting about religious topics also trying to avoid swearing; this hypothesis can also be supported with less strong swearing alternatives ('crap', 'freaking', 'dang') prevailing among the negatively weighed words along the religious words.

If the native language of a group is different from that of the majority, the words of this different language also stand out from the overall structure, as there is naturally a stronger correlation among words belonging to the same language. Therefore the applied method can discover languages different from that of the bulk of the sample. In the sixth singular vector, we can observe this phenomenon with Spanish words, which form more than the third of the positively weighed wordcloud (Fig.~\ref{fig:svect}j.). The English terms 'Mexico' and 'Mexican' also appear in this group, which shows that concepts related to the topic are also identified even if they do not belong to the discovered language.

\begin{figure*}[h]
\centering
\includegraphics[width=.7\textwidth]{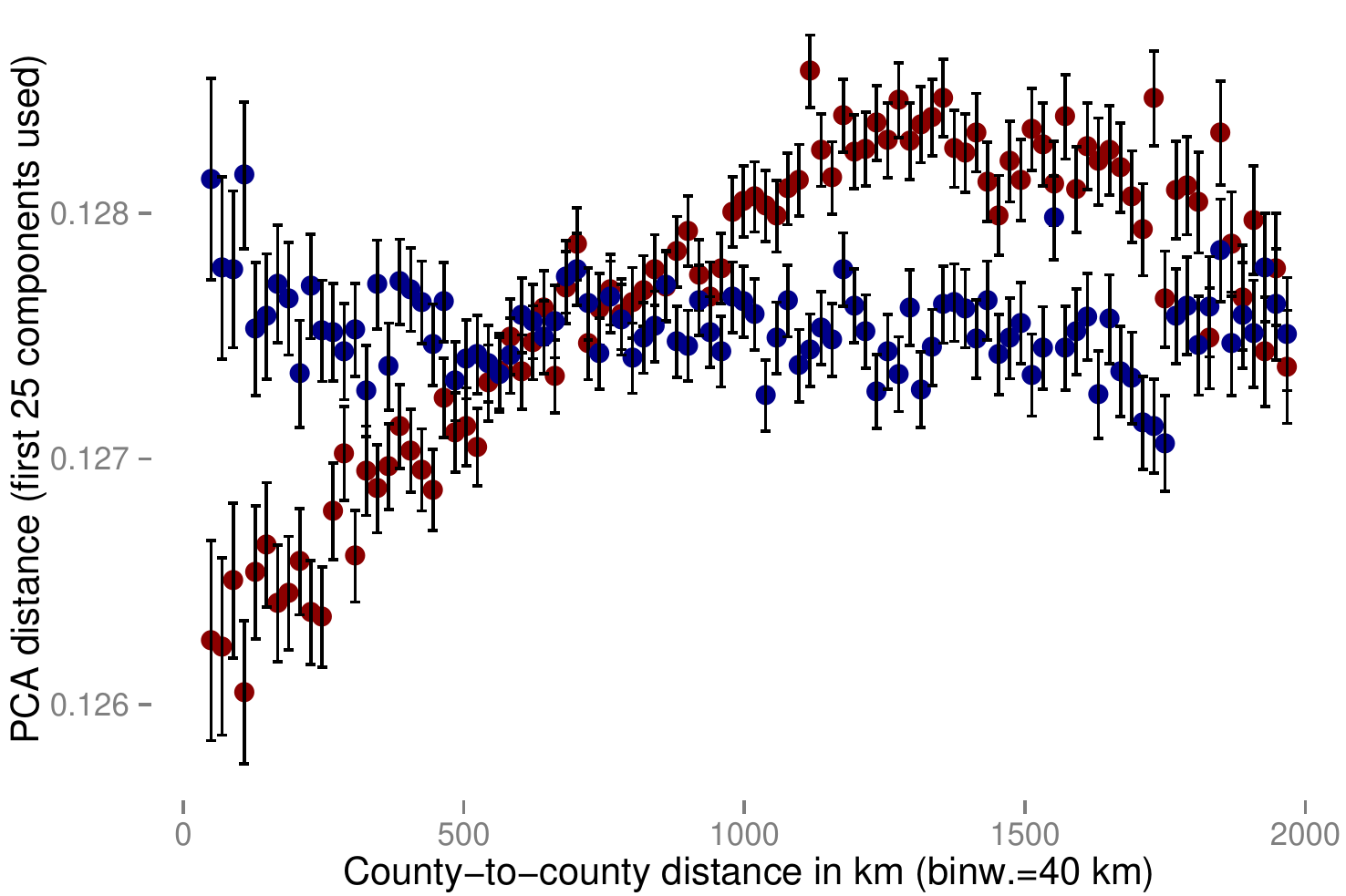}
\caption{\textbf{Semantic versus real-world distance of counties.} Euclidean distance of counties in the semantic subspace of the first 25 components obtained from LSA as a function of geographical distance (red dots). Baseline calculated from a random permutation of counties (blue dots). Errorbars correspond to the error of the binwise means.}
\label{fig:dist}
\end{figure*}

\begin{figure*}[h]
	\centerline{\includegraphics[height=0.85\textheight]{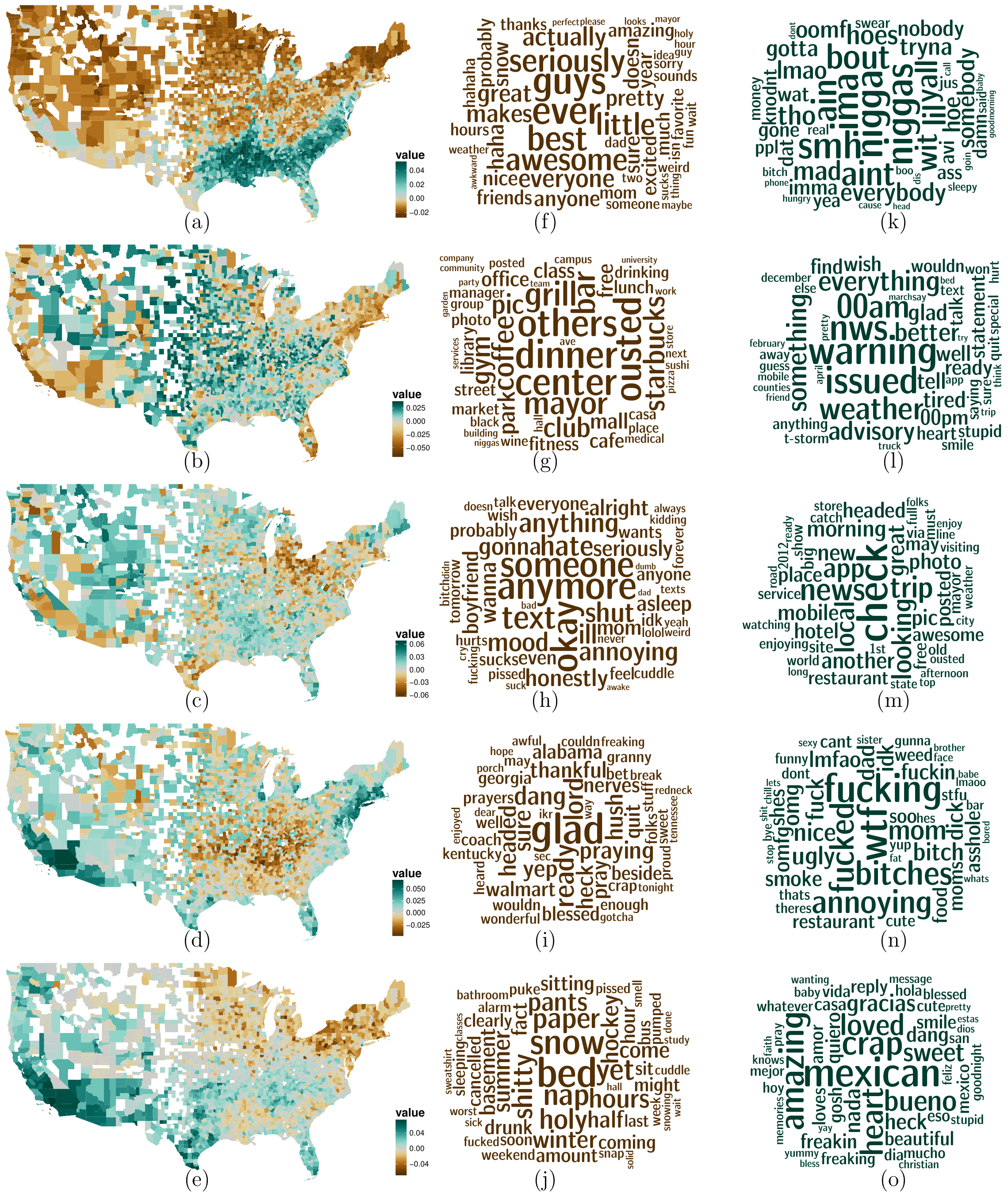}}
	\caption{\textbf{Representations of the first few right (a-e) and left (f-o) singular vectors.} Weights corresponding to counties are plotted on a US map, brown representing the negatively, blue the positively weighed counties. Wordclouds represent left singular vectors with coloring corresponding to that of the maps, and word size representing the weight of each word in each actual singular vector.}
	\label{fig:svect}
\end{figure*}

Similarly to topic identification, where semantically close words form topics, analyzing regional patterns reveal documents that are close to each other in the semantic space spanned by these topics. Plotting the right singular vectors on a map (Fig. \ref{fig:svect}a-e.), the most striking feature is the regional proximity of documents having close weights in the singular vectors. Document-by-document (county-by-county) Euclidean distances in the PCA subspace of the first 25 component as the function of real county-by-county centroid distances \footnote{\url{http://cta.ornl.gov/transnet/SkimTree.htm}} illustrate this observation. In Fig.~\ref{fig:dist}. mean PCA subspace distances (red dots) are plotted for each 40~km range of real county centroid distances. As a baseline, the same is done for a random permutation of counties (blue dots). It is remarkable that below 500~km, counties are closer in the semantic space, as could be expected from a random realization. From 700~km to 1800~km, semantic distance is greater than it would be randomly. Geographical proximity is thus a main driving force in the similarity of language patterns in Twitter-space.

Analyzing these geographical patterns in each singular vectors provides insights into the regional distribution of
the single topics. On a US map, the second component (Fig.~\ref{fig:svect}a.), which is responsible for the most variance in the Twitter data, emerges as a block in the Southeastern part of the US. Apart from the big Southeastern block, Chicago and Detroit are also marked by this pattern of language usage. In the third component (Fig.~\ref{fig:svect}b.), negative weights (brown patches) mark the biggest cities and surrounding counties which belong to their agglomeration. The most positive pattern of the fourth component (Fig.~\ref{fig:svect}c.) reveals some important touristic attractions such as the center of New York, Washington and San Francisco, the Craters of the Moon National Monument and Preserve in Idaho, Aspen Mountain ski area in Colorado or Hawaii. The regional pattern of the fifth component (Fig.~\ref{fig:svect}d.) is less obvious, though a part of the central US and the Southeastern block is discernible in the religion-related end of the component. The sixth component distinguishes the Southwestern part and the Northwestern corner of the US (Fig.~\ref{fig:svect}e.), Florida and some bigger cities such as New York or Chicago.

\begin{figure*}[h]
\centerline{\includegraphics[height=0.85\textheight]{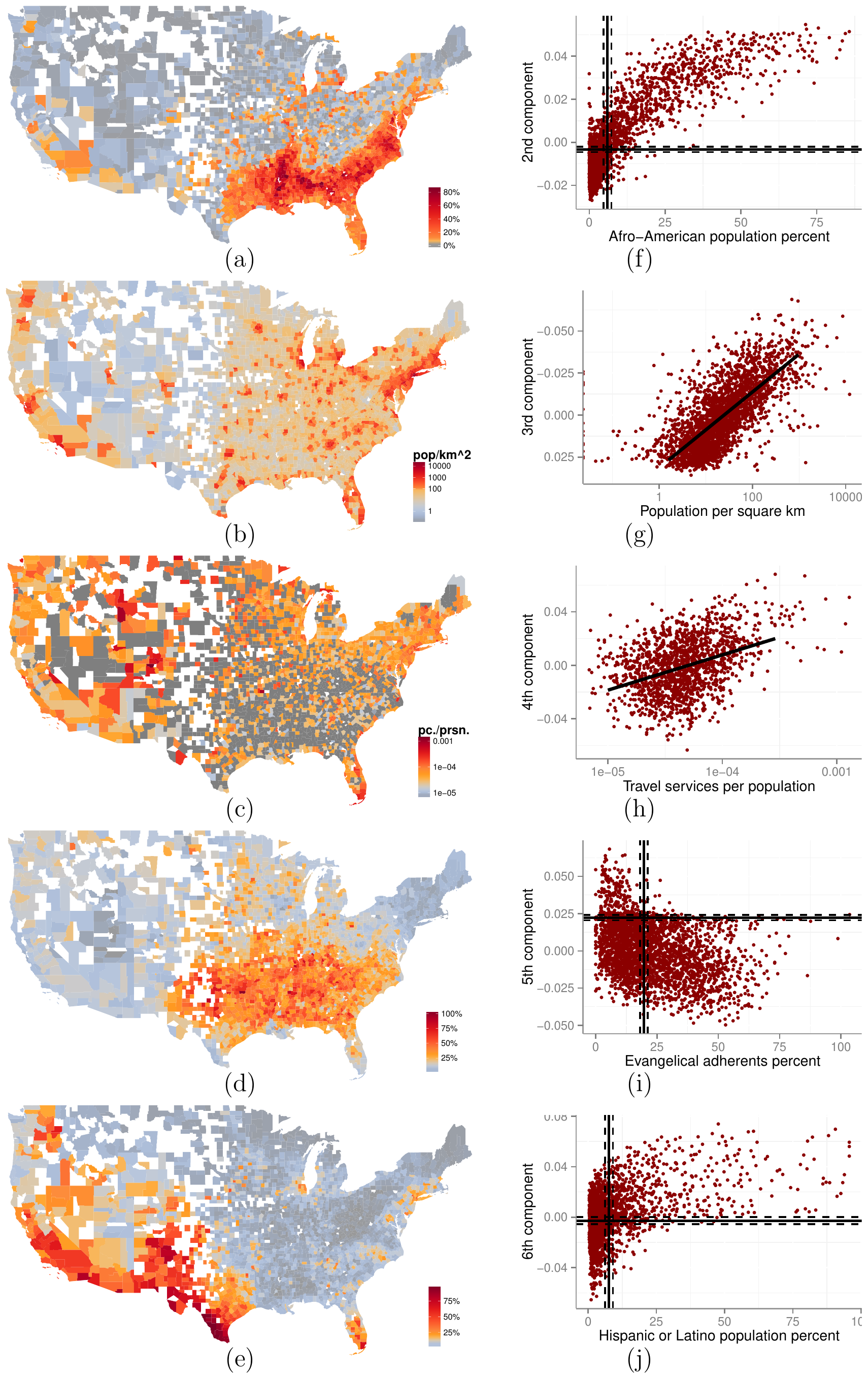}}
\caption{\textbf{Census data with most correlation for each component displayed on a map (a-e) and scatterplots of these data series with county weights from corresponding singular vectors (f-j).} Lines representing a symmetric relationship (Pearson correlation) are drawn on scatterplots (g) and (h). On scatterplots (f), (i) and (j), horizontal and vertical lines correspond to the best segmentation when testing for a Boolean relationship between variables. The points between the dashed lines were not taken into account when calculating the test statistics for the sparseness of each segment.}
\label{fig:sdemo}
\end{figure*}

To discover possible governing factors of the geographical language variation patterns and their relation to demography, we calculate Pearson correlation values between right singular vectors and data obtained from the US Census Bureau described in Section~2.3. Data series that have the greatest absolute correlation values (p$<$0.0001, Bonferroni-corrected) with each component are shown in Table~1. The large correlation (0.872) of the second component with the population proportion of African-Americans per county indicates that the observed slang words and the blockwise regional pattern are linked to the presence of this demographic group (note that, however, we have no evidence of whether the tweets causing the variation were indeed posted by African-American people). Fig.~\ref{fig:sdemo}a. shows the census proportions on a US map, with the regional pattern approximately corresponding to that of the singular vector. It is with noting that apart from the large Southeastern block, Chicago and Detroit are also marked by having the characteristic slang word pattern, as well as a higher proportion of African-American population. A similarly large correlation (0.500) with ethnicity (Hispanic or Latino origin) also arises is the case of the sixth component, as expected from the observed Spanish words and the Southwestern positive weights on the map. Fig.~\ref{fig:sdemo}. shows the percent of people with Hispanic or Latino origin in US counties, the distribution resembling that of the right singular vector.

The data series that show the largest correlation with the third component are resident total population rank (0.844) and rural-urban continuum code \footnote{\url{http://www.ers.usda.gov/data-products/rural-urban-continuum-codes.aspx}} (0.630). Since neither are continuous variables, we instead show population density values in each county on the map of Fig.~\ref{fig:sdemo}b. Densely populated areas mark the biggest cities and their surrounding agglomerations of the US, and these areas are also discernible in the brown patches of the third singular vector in Fig.~\ref{fig:svect}b. It confirms the idea of the most densely populated areas giving the negative end of the third singular vector in both the words and their regional distribution. A basic feature of the Twitter corpus is thus linked simply to city lifestyle, more generally to the associated socioeconomic status.

Correlation values show whether there exists some relation between the language patterns and demographic data (see Table ~1). Analyzing scatterplots of the greatest correlations provides us some insight into the structure of these relations. Plotting the regional weights of the second and sixth singular vector against African-American and Hispanic or Latino ethnicity percentages exhibits very similar features (Fig.~3f,3j). Correlation analysis also revealed that a prevalence of evangelical religious groups (Baptists and Methodists) is related to (-0.372) the religious content of the fifth component (see Fig.~3i); county-level rates of adherence of evangelical churches are plotted in Fig.~\ref{fig:sdemo}d. The existence of a virtual 'Bible Belt' is thus confirmed in the Twitter-space, corresponding to former identification of religious groups in cyberspaces~\cite{Shelton_2012,Zook_2010}. An opposite correlation is present with Catholic and Orthodox churches, which we speculate to be the consequence of these having a smaller attendance in counties where evangelical churches are more prominent.

Although almost all of the above-described correlations could be explained by an underlying function, a boolean implication model description seems more plausible. Boolean implications have already been used in gene expression research~\cite{Sahoo_2008}, to uncover non-symmetric relationships where correlation analysis would only partially or not at all measure connection between two variables. In the case of ethnicities, if we take $y$ values as a measure of how strongly slang (Fig.~3f) or Spanish (Fig.~3j) (see the wordclouds of Fig.~2f and Fig.~2j) is present in the Twitter messages of the counties, we can observe that below a certain ratio of ethnicity prevalence (6.0\% in the second component and 7.6\% in the sixth), language patterns show different levels of non-slang or non-Spanish usage. If ethnicity prevalence is greater than the threshold value, slang or Spanish usage rises steeply with growing ethnicity proportion. Above the threshold, there are very few counties with non-slang or non-Spanish language patterns. In this terminology, the two scatterplots corresponding to ethnicity prevalence can be translated to 'high ethnicity rates $\Rightarrow$ missing non-slang/non-Spanish' words implication. The limits corresponding to the best implication model were the mentioned $5.99\%\pm1.28\%$ and $7.65\%\pm1.43\%$ of prevalence for the two ethnic groups, with $-0.00328\pm0.00123$ and $-0, 00277\pm0.00209$ as a limit on the axes of the second and sixth component. The measures of sparseness for the lower right segments were $\delta=21.941$, $\epsilon=0.036$, and $\delta=12.98$, $\epsilon=0.15$, respectively.

A boolean implication also describes the scatterplot of the fifth component measured against evangelical adherence rates (Fig.~3i). Here the $y$ axis represents a level of swearing (cf.~the words in Fig.~\ref{fig:svect}i) present in the Twitter-sphere of tweets posted in a county. Thus the implication can be translated to 'high evangelical prevalence rates $\Rightarrow$ low swearing level'. The pattern implies a stronger connection between the two variables, as could be inferred from the symmetric correlation measure. It seems as if above a certain adherence rate, a text with a high swearing level could not propagate further or could not find way to broader discussion. Here the automatically detected limit was at a $19.67\%\pm1.55\%$ adherence level, the limit on the 'swearing' axis lies at $0.02230\pm0.00177$, and the lower right corner showed a significantly large sparseness with $\delta=8.579$ and $\epsilon=0.013$.

\section{Discussion}

We can conclude that the applied unsupervised learning method successfully discovers topics and their regional patterns in the Twitter-sphere, with county weights in right singular vectors representing a distance in the semantic space along a topic given by word weights of the left singular vectors. It is also remarkable that geographical closeness implies closeness in the semantic space, which suggests that language usage is on a certain level bound to geographical proximity.

We also find that regional patterns in language use are driven not just by geographical proximity, but socioeconomical and cultural similarities, like degree of urbanization, religion or ethnicity. It seems that the most important factor behind the variation in the language use of different counties is the presence of Afro-American ethnicity, as confirmed by the significant correlation between the census-based share of Afro-American population and the appropriate county weights. Corresponding word weights mirror this observation with words representative of the typical slang use associated with this ethnicity. This type of slang use thus turns out to be the most distinguishing factor in everyday US Twitter conversation.

Following ethnicity, the second most important feature found in Twitter language is related to the population density of a county. The  interpretation could be that beyond ethnicity, our everyday language is largely influenced by our surroundings. Thus living in densely populated places, which means mostly living in urban areas, results in words specific to urban lifestyle appearing more frequently in user messages.

The language footprints of tourism can also be captured by our method, suggesting that the effect of messages or users being on a holiday should always be considered, when trying to relate online content to real-world phenomena.

Some relations are better described by a non-symmetric boolean implication model instead of the symmetric correlation measure. We find that the presence of ethnic groups above a certain threshold implies a weight greater than a certain level along the semantic axis corresponding to the component connected to this ethnic group. We also find that counties exhibiting high evangelical adherence rates show low level on the ’swearing scale’ given by the corresponding component. This is interesting, since the phenomenon cannot be observed with the two other major denominations, the Catholic and Orthodox churches. It suggests that the online presence of Evangelical churches is inherently different from that of the other denominations, and its adherents have a significant effect on the word choice on the Twitter platform.

Our results suggest that online social network activity can be used effectively to monitor the spatial variation of cultural traits as represented in language use, yielding an up-to-date picture of important social phenomena. We believe our present study demonstrates an approach for measuring the importance of certain demographic attitudes when working with textual Twitter data. We suggest, therefore, that it could form the basis of further research focusing on the evaluation of demographic data estimation from other sources, or on the dynamical processes that result in the patterns found here. While our results were obtained using the Twitter microblogging platform, research could be further extended to investigate whether the incorporation of other metadata (e.g. user activity, user mobility, user profile descriptions etc) or the analysis of different text sources could refine or enhance our findings.

\section*{Acknowledgements}
The authors would like to thank the partial support of the European Union and the
European Social Fund through the FuturICT.hu project (Grant No.: TAMOP-4.2.2.C-
11/1/KONV-2012-0013), the OTKA-103244, OTKA-114560, Ericsson and the MAKOG
Foundation.

\nocite{*}

%\bibliography{biblio2}

\begin{thebibliography}{10}

\bibitem{Brain2005}
D.~Brain.
\newblock {From Good Neighborhoods to Sustainable Cities: Social Science and
  the Social Agenda of the New Urbanism}.
\newblock {\em International Regional Science Review}, 28(2):217--238, 2005.

\bibitem{Quillian1999}
Lincoln Quillian.
\newblock {Migration Patterns and the Growth of High- Poverty Neighborhoods ,
  1970 -- 1990 1}.
\newblock {\em American Journal of Sociology}, 105(1):1--37, 1999.

\bibitem{Sampson2009}
Robert~J Sampson.
\newblock {Disparity and diversity in the contemporary city: social (dis)order
  revisited}.
\newblock {\em The British Journal of Sociology}, 60(1):1--31, March 2009.

\bibitem{Iceland2006}
John Iceland and Rima Wilkes.
\newblock {Does Socioeconomic Status Matter? Race, Class, and Residential
  Segregation}.
\newblock {\em Social Problems}, 53(2):248--273, May 2006.

\bibitem{Bruch2006}
Elizabeth~E. Bruch and Robert~D. Mare.
\newblock {Neighborhood Choice and Neighborhood Change}.
\newblock {\em American Journal of Sociology}, 112(3):667--709, November 2006.

\bibitem{Bettencourt2007}
Lu{\'i}s M~a Bettencourt, Jos{\'e} Lobo, Dirk Helbing, Christian K{\"u}hnert,
  and Geoffrey~B West.
\newblock {Growth, innovation, scaling, and the pace of life in cities.}
\newblock {\em Proceedings of the National Academy of Sciences of the United
  States of America}, 104(17):7301--7306, 2007.

\bibitem{Cummings2012}
David Cummings, Haruki Oh, and Ningxuan Wang.
\newblock {Who Needs Polls? Gauging Public Opinion from Twitter Data}.
\newblock 2012.

\bibitem{OConnor2010}
Brendan O{\rq}Connor, Ramnath Balasubramanyan, Bryan~R Routledge, and Noah~a
  Smith.
\newblock {From tweets to polls: Linking text sentiment to public opinion time
  series}.
\newblock In {\em {ICWSM}}, volume~11, pages 1--2, 2010.

\bibitem{Deville}
Pierre Deville, Catherine Linard, Samuel Martin, Marius Gilbert, Forrest~R
  Stevens, and Andrea~E Gaughan.
\newblock {Dynamic population mapping using mobile phone data}.
\newblock {\em Proceedings of the National Academy of Sciences},
  111(45):15888--15893, 2014.

\bibitem{Botta2015}
Federico Botta, Helen~Susannah Moat, Tobias Preis, Moat Hs, and Tobias Preis.
\newblock {Quantifying crowd size with mobile phone and Twitter data}.
\newblock {\em Royal Society Open Science}, 2(5):150162, 2015.

\bibitem{Louail}
Thomas Louail, Maxime Lenormand, Oliva~G Cant{\'u}-Ros, Miguel Picornell,
  Ricardo Herranz, Enrique Frias-Martinez, Jos{\'e}~J Ramasco, and Marc
  Barth{\'e}lemy.
\newblock {From mobile phone data to the spatial structure of cities.}
\newblock {\em Scientific reports}, 4:5276, jan 2014.

\bibitem{Frias}
Vanessa Frias-Martinez and Enrique Frias-Martinez.
\newblock {Spectral clustering for sensing urban land use using Twitter
  activity}.
\newblock {\em Engineering Applications of Artificial Intelligence},
  35(10):237--245, 2014.

\bibitem{Ratti2007cellularcensus}
Jonathan Reades, Francesco Calabrese, Andres Sevtsuk, and Carlo Ratti.
\newblock {Cellular census: Explorations in urban data collection}.
\newblock {\em Pervasive Computing, IEEE}, 6(3):30--38, 2007.

\bibitem{Krings}
Vincent~D Blondel, Adeline Decuyper, and Gautier Krings.
\newblock {A survey of results on mobile phone datasets analysis}.
\newblock {\em EPJ Data Science}, 4(1):10, 2015.

\bibitem{mislove}
Alan Mislove.
\newblock {\em {Online Social Networks: Measurement, Analysis, and Applications
  to Distributed Information Systems}}.
\newblock PhD thesis, Rice University, 2009.

\bibitem{Cheng2011}
Zhiyuan Cheng, James Caverlee, Kyumin Lee, and Daniel~Z. Sui.
\newblock {Exploring Millions of Footprints in Location Sharing Services}.
\newblock In {\em {International AAAI Conference on Web and Social Media}},
  pages 81--88, 2011.

\bibitem{george}
D~Brockmann, L~Hufnagel, and T~Geisel.
\newblock {The scaling laws of human travel.}
\newblock {\em Nature}, 439(7075):462--5, jan 2006.

\bibitem{george2}
Christian Thiemann, Fabian Theis, Daniel Grady, Rafael Brune, and Dirk
  Brockmann.
\newblock {The structure of borders in a small world.}
\newblock {\em PloS one}, 5(11):e15422, jan 2010.

\bibitem{bbva}
Stanislav Sobolevsky, Izabela Sitko, Remi Tachet, Juan~Murillo Arias, and Carlo
  Ratti.
\newblock {Cities through the Prism of People{\rq}s Spending Behavior}.
\newblock {\em PLoS ONE}, 11(2):e0146291, 2016.

\bibitem{cargps}
Luca Pappalardo, Salvatore Rinzivillo, Zehui Qu, Dino Pedreschi, and Fosca
  Giannotti.
\newblock {Understanding the patterns of car travel}.
\newblock {\em The European Physical Journal Special Topics}, 215(1):61--73,
  jan 2013.

\bibitem{returners}
Luca Pappalardo, Filippo Simini, Salvatore Rinzivillo, Dino Pedreschi, Fosca
  Giannotti, and Albert-L{\'a}szl{\'o} Barab{\'a}si.
\newblock {Returners and explorers dichotomy in human mobility}.
\newblock {\em Nature Communications}, 6:8166, 2015.

\bibitem{oyster}
Camille Roth, Soong~Moon Kang, Michael Batty, and Marc Barth{\'e}lemy.
\newblock {Structure of Urban Movements: Polycentric Activity and Entangled
  Hierarchical Flows}.
\newblock {\em PLoS ONE}, 6(1):e15923, 01 2011.

\bibitem{oyster2}
Samiul Hasan, Christian Schneider, Satish Ukkusuri, and Marta Gonz{\'a}lez.
\newblock {Spatiotemporal Patterns of Urban Human Mobility.}
\newblock {\em Journal of Statistical Physics}, 151(1/2):304--318, 2013.

\bibitem{Sloan2015}
Luke Sloan, Jeffrey Morgan, Pete Burnap, and Matthew Williams.
\newblock {Who Tweets? Deriving the Demographic Characteristics of Age,
  Occupation and Social Class from Twitter User Meta-Data}.
\newblock {\em Plos One}, 10(3):e0115545, 2015.

\bibitem{Longley2015}
Paul~a Longley, Muhammad Adnan, and Guy Lansley.
\newblock {The geotemporal demographics of Twitter usage}.
\newblock {\em Environment and Planning A}, 47(2):465--484, 2015.

\bibitem{delineating}
Stanislav Sobolevsky, Michael Szell, Riccardo Campari, Thomas Couronn{\'e},
  Zbigniew Smoreda, and Carlo Ratti.
\newblock {Delineating geographical regions with networks of human interactions
  in an extensive set of countries}.
\newblock {\em PLoS ONE}, 8(12):e81707, 2013.

\bibitem{spatialfingerprints}
Zs{\'o}fia Kallus, Norbert Barankai, J{\'a}nos Sz{\"u}le, and G{\'a}bor Vattay.
\newblock {Spatial Fingerprints of Community Structure in Human Interaction
  Network for an Extensive Set of Large-Scale Regions}.
\newblock {\em Plos One}, 10(5):e0126713, 2015.

\bibitem{Mocanu2013}
Delia Mocanu, Andrea Baronchelli, Nicola Perra, Bruno Gon\c{c}alves, Qian
  Zhang, and Alessandro Vespignani.
\newblock {The Twitter of Babel: Mapping World Languages through Microblogging
  Platforms}.
\newblock {\em PLoS ONE}, 8(4):e61981, 2013.

\bibitem{Grauwin}
S{\'e}bastian Grauwin, Stanislav Sobolevsky, Simon Moritz, Istv{\'a}n
  G{\'o}dor, and Carlo Ratti.
\newblock {\em {Towards a comparative science of cities: using mobile traffic
  records in New York, London and Hong Kong}}, volume~13 of {\em
  {Geotechnologies and the Environment}}, pages 363--387.
\newblock 2014.

\bibitem{Gonzalez2008mobility}
Marta~C Gonz{\'a}lez, C{\'e}sar~A Hidalgo, and Albert-L{\'a}szl{\'o}
  Barab{\'a}si.
\newblock {Understanding individual human mobility patterns.}
\newblock {\em Nature}, 453(7196):779--82, jun 2008.

\bibitem{Jiang}
Shan Jiang, Joseph Ferreira, and Marta~C Gonz{\'a}lez.
\newblock {Activity-Based Human Mobility Patterns Inferred from Mobile Phone
  Data: A Case Study of Singapore}.
\newblock In {\em {Int. Workshop on Urban Computing}}, 2015.

\bibitem{gowalla}
Eunjoon Cho, SA~Myers, and Jure Leskovec.
\newblock {Friendship and mobility: user movement in location-based social
  networks}.
\newblock In {\em {Proceedings of the 17th ACM SIGKDD International Conference
  on Knowledge Discovery and Data Mining}}, pages 1082--1090, 2011.

\bibitem{Hawelka2014}
Bartosz Hawelka, Izabela Sitko, Euro Beinat, Stanislav Sobolevsky, Pavlos
  Kazakopoulos, and Carlo Ratti.
\newblock {Geo-located Twitter as proxy for global mobility patterns}.
\newblock {\em Cartography and Geographic Information Science}, 41(3):260--271,
  feb 2014.

\bibitem{Simini}
Filippo Simini, Marta~C. Gonz{\'a}lez, Amos Maritan, and Albert-L{\'a}szl{\'o}
  Barab{\'a}si.
\newblock {A universal model for mobility and migration patterns}.
\newblock {\em Nature}, 484(7392):96--100, 2012.

\bibitem{Landauer1997}
TK~Landauer and ST~Dumais.
\newblock {A solution to Plato's problem: The latent semantic analysis theory
  of acquisition, induction, and representation of knowledge.}
\newblock {\em Psychological review}, 1(2):211--240, 1997.

\bibitem{Deerwester1990}
SC~Deerwester, ST~Dumais, and TK~Landauer.
\newblock {Indexing by latent semantic analysis}.
\newblock {\em Journal of the American society for Information Science},
  41(6164):391, 1990.

\bibitem{Schwartz2013a}
H~Andrew Schwartz, Johannes~C Eichstaedt, Margaret~L Kern, Lukasz Dziurzynski,
  Stephanie~M Ramones, Megha Agrawal, Achal Shah, Michal Kosinski, David
  Stillwell, Martin E~P Seligman, and Lyle~H Ungar.
\newblock {Personality, gender, and age in the language of social media: the
  open-vocabulary approach.}
\newblock {\em PloS one}, 8(9):e73791, jan 2013.

\bibitem{Petersen2012}
Alexander~M. Petersen, Joel~N. Tenenbaum, Shlomo Havlin, H.~Eugene Stanley, and
  Matja\v{z} Perc.
\newblock {Languages cool as they expand: Allometric scaling and the decreasing
  need for new words}.
\newblock {\em Scientific Reports}, 2:943, 2012.

\bibitem{Perc2012}
M.~Perc.
\newblock {Evolution of the most common English words and phrases over the
  centuries}.
\newblock {\em Journal of The Royal Society Interface}, 9(July):3323--3328,
  2012.

\bibitem{Eichstaedt2015}
Johannes~C Eichstaedt, Hansen~Andrew Schwartz, Margaret~L Kern, Gregory Park,
  Darwin~R Labarthe, Raina~M Merchant, Sneha Jha, Megha Agrawal, Lukasz~a
  Dziurzynski, Maarten Sap, Christopher Weeg, Emily~E Larson, Lyle~H Ungar, and
  Martin E~P Seligman.
\newblock {Psychological Language on Twitter Predicts County-Level Heart
  Disease Mortality}.
\newblock {\em Psychological Science}, 26(2):159--169, 2015.

\bibitem{Seresinhe2015}
Chanuki~Illushka Seresinhe, Tobias Preis, and Helen~Susannah Moat.
\newblock {Quantifying the Impact of Scenic Environments on Health}.
\newblock {\em Scientific Reports}, 5:16899, 2015.

\bibitem{unemployment}
Alejandro Llorente, Manuel Cebrian, and Esteban Moro.
\newblock {Social media fingerprints of unemployment}.
\newblock {\em PLoS ONE}, 10(5):e0128692, 2015.

\bibitem{Pavlicek2015}
Jaroslav Pavlicek and Ladislav Kristoufek.
\newblock {Nowcasting Unemployment Rates with Google Searches: Evidence from
  the Visegrad Group Countries}.
\newblock {\em Plos One}, 10(5):e0127084, 2015.

\bibitem{Curme2014}
C.~Curme, T.~Preis, H.~E. Stanley, and H.~S. Moat.
\newblock {Quantifying the semantics of search behavior before stock market
  moves}.
\newblock {\em Proceedings of the National Academy of Sciences},
  111(32):11600--11605, 2014.

\bibitem{Cheng2010}
Zhiyuan Cheng, James Caverlee, and Kyumin Lee.
\newblock {You are where you tweet: a content-based approach to geo-locating
  twitter users}.
\newblock In {\em {Proceedings of the 19th ACM International Conference on
  Information and Knowledge Management}}, pages 759--768, 2010.

\bibitem{Backstrom2010}
Lars Backstrom, Eric Sun, and Cameron Marlow.
\newblock {Find me if you can: improving geographical prediction with social
  and spatial proximity}.
\newblock In {\em {Proceedings of the 19th international conference on World
  wide web}}, pages 61--70. ACM, 2010.

\bibitem{travelingtrends}
Emilio Ferrara, Onur Varol, Filippo Menczer, and Alessandro Flammini.
\newblock {Traveling trends: social butterflies or frequent fliers?}
\newblock In {\em {COSN '13 Proceedings of the first ACM conference on Online
  social networks}}, pages 213--222, 2013.

\bibitem{Eisenstein2012}
Jacob Eisenstein, Brendan O'Connor, Noah~a. Smith, and Eric~P. Xing.
\newblock {Diffusion of Lexical Change in Social Media}.
\newblock {\em PLoS ONE}, 9(11):e113114, 11 2014.

\bibitem{Mitchell2013}
Lewis Mitchell, Morgan~R Frank, Kameron~Decker Harris, Peter~Sheridan Dodds,
  and Christopher~M Danforth.
\newblock {The geography of happiness: connecting twitter sentiment and
  expression, demographics, and objective characteristics of place.}
\newblock {\em PloS one}, 8(5):e64417, jan 2013.

\bibitem{Lin2010}
Zhouchen Lin, Minming Chen, and Yi~Ma.
\newblock {The Augmented Lagrange Multiplier Method for Exact Recovery of
  Corrupted Low-Rank Matrices}.
\newblock 2010.

\bibitem{Candes2011}
EJ~Cand{\`e}s, Xiaodong Li, Y~Ma, and John Wright.
\newblock {Robust principal component analysis?}
\newblock {\em Journal of the ACM}, 58(3):11, 2011.

\bibitem{Morstatter2013}
F~Morstatter, J~Pfeffer, H~Liu, and K~Carley.
\newblock {Is the Sample Good Enough ? Comparing Data from Twitter {\rq} s
  Streaming API with Twitter {\rq} s Firehose}.
\newblock In {\em {International Conference on Weblogs and Social Media}},
  pages 400--408, 2013.

\bibitem{Dobos2013}
Laszlo Dobos, Janos Szule, Tamas Bodnar, Tamas Hanyecz, Tamas Sebok, Daniel
  Kondor, Zsofia Kallus, Jozsef Steger, Istvan Csabai, and Gabor Vattay.
\newblock {A multi-terabyte relational database for geo-tagged social network
  data}.
\newblock In {\em {4th IEEE International Conference on Cognitive
  Infocommunications, CogInfoCom 2013 - Proceedings}}, pages 289--294, 2013.

\bibitem{Szalay2007}
AS~Szalay, Jim Gray, George Fekete, and PZ~Kunszt.
\newblock {Indexing the sphere with the hierarchical triangular mesh}.
\newblock {\em arXiv}, (arXiv:cs/0701164), 2007.

\bibitem{geomhtm}
D{\'a}niel Kondor, L{\'a}szl{\'o} Dobos, Istv{\'a}n Csabai, Andr{\'a}s Bodor,
  G{\'a}bor Vattay, Tam{\'a}s Budav{\'a}ri, and Alexander~S. Szalay.
\newblock {Efficient classification of billions of points into complex
  geographic regions using hierarchical triangular mesh}.
\newblock In {\em {Proceedings of the 26th International Conference on
  Scientific and Statistical Database Management - SSDBM '14}}, pages 1--4, New
  York, New York, USA, 2014. ACM Press.

\bibitem{Gotoh1997}
Y~Gotoh and S~Renals.
\newblock {Document space models using latent semantic analysis.}
\newblock In {\em {Proc. Eurospeech}}, pages 1443--1446, 1997.

\bibitem{Sahoo_2008}
Debashis Sahoo, David~L Dill, Andrew~J Gentles, Robert Tibshirani, and Sylvia~K
  Plevritis.
\newblock {Boolean implication networks derived from large scale whole genome
  microarray datasets}.
\newblock {\em Genome Biology}, 9(10):R157, 2008.

\bibitem{Shelton_2012}
Taylor Shelton, Matthew Zook, and Mark Graham.
\newblock {The Technology of Religion: Mapping Religious Cyberscapes}.
\newblock {\em The Professional Geographer}, 64(4):602--617, nov 2012.

\bibitem{Zook_2010}
Matthew Zook and Mark Graham.
\newblock {Featured graphic: The virtual `bible belt'}.
\newblock {\em Environ. Plann. A}, 42(4):763--764, 2010.

\end{thebibliography}
%\bibliographystyle{unsrt}

\section{Data availability statement}

Owing to Twitter’s policy we cannot publicly share the original dataset used in this
analysis. The county-wide word frequency matrix and the results of the LSA compiled
are available in the Dataverse repository at \url{http://dx.doi.org/10.7910/
DVN/EXWJRJ} and also at \url{http://www.vo.elte.hu/papers/2016/twitter-pca}.

\newpage

\clearpage

\begin{table*}
\begin{tabular*}{\hsize}{@{\extracolsep{\fill}}crl}
Comp. no. & $\rho$ & Dataset\\ \hline \hline
2 & 0.87  & Population of one race - percent Black or African American alone (2010)                       \\
  & 0.77  & Owner-occupied housing units, African American householder, per population (2010)   \\
  & -0.75 & Population of one race - percent White alone (2010)                                           \\
  & 0.65  & Black Protestant - Rates of adherence per 1,000 population (2010)                                             \\\hline
3 & 0.84  & Resident total population estimate - rank (2007)															\\
  & -0.72 & Population density (2010)                                                                                           \\
  & -0.72 & Percent of adults with a bachelor's degree or higher (2008-2012)                                              \\
  & 0.63  & Rural-urban Continuum Code (2013)                                                                             \\\hline
4 & 0.28  & All Other Travel Arrangement and Reservation Services Total Number of Establishments                         \\
  & 0.28  & Tour Operators Total Number of Establishments                                                                \\
  & 0.26  & Convention and Visitors Bureaus Total Number of Establishments                                               \\
  & 0.26  & Accommodation establishments with payroll per population (2007) \\\hline
5 & 0.39  & Catholic - Rates of adherence per 1,000 population (2010)                                                     \\
  & -0.37 & Evangelical Protestant - Rates of adherence per 1,000 population (2010)                                       \\
  & 0.36  & Orthodox - Total number of adherents (2010)                                                                   \\
  & -0.29 & Evangelical Protestant - Total number of adherents per population (2010)                                      \\ \hline
6 & 0.50  & Percent Hispanic or Latino population (2010)                                                  \\
  & 0.50  & Hispanic or Latino population - Percent Mexican (2010)                                                         \\
  & 0.38  & Average household size (2010)                                                                                  \\
  & 0.36  & Percent households with persons under 18 years (2010)                                                   
\end{tabular*}
\caption{\textbf{Correlations with demographic data series.} Greatest Pearson-correlations (p$<$0.0001 at a Bonferroni-corrected level) in the demographic datasets with the first few singular vectors}
\end{table*}

\end{document}